\documentclass{article}

\usepackage{PRIMEarxiv}

\usepackage[utf8]{inputenc} 
\usepackage[T1]{fontenc}    
\usepackage{hyperref}       
\usepackage{url}            
\usepackage{booktabs}       
\usepackage{amsfonts}       
\usepackage{nicefrac}       
\usepackage{microtype}      
\usepackage{lipsum}
\usepackage{fancyhdr}       
\usepackage{graphicx}       
\usepackage{siunitx}
\usepackage[numbers]{natbib}
\usepackage{todonotes}
\usepackage{subcaption}
\usepackage{tikz,xcolor,hyperref}
\graphicspath{{media/}}     

\DeclareRobustCommand{\orcidicon}{%
    \begin{tikzpicture}
    \draw[lime, fill=lime] (0,0) 
    circle [radius=0.16] 
    node[white] {{\fontfamily{qag}\selectfont \tiny ID}};    \draw[white, fill=white] (-0.0625,0.095) 
    circle [radius=0.007];    \end{tikzpicture}
    \hspace{-2mm}}
\foreach \x in {A, ..., Z}{%
    \expandafter\xdef\csname orcid\x\endcsname{\noexpand\href{https://orcid.org/\csname orcidauthor\x\endcsname}{\noexpand\orcidicon}}
    }
\title{Coarse-graining bistability with the Martini force field
\thanks{\textit{\underline{Citation}}: 
\textbf{Alexander D. Muratov, Vladik A. Avetisov;} Coarse-graining bistability with the Martini force field. \textit{Physics of Fluids} 1 December 2024; 36 (12): 127157 \href{https://doi.org/10.1063/5.0246426}{DOI:10.1063/5.0246426}.} 
}

\author{
  Alexander D. Muratov\orcidA, Vladik A. Avetisov\orcidB \\
  Semenov Federal Research Center for Chemical Physics, Russian Academy of Sciences, 119991 Moscow, Russia\\
  Design Center for Molecular Machines, Moscow, Russia \\
  \texttt{alexander.muratov@chph.ras.ru~(A.D.M.); avetisov@chph.ras.ru~(V.A.A.);} \\
}

\begin{document}
\maketitle

\begin{abstract}
An increasing interest in molecular structures whose long term dynamics resemble those of bistable mechanical systems promotes the search of possible candidates that may operate as two-state switching units. Of particular interest are the systems that are capable to exhibit such bistable effects as spontaneous vibrations, stochastic resonance and spontaneous synchronization. Previously, in all-atom molecular dynamics simulations it has been demonstrated that short pyridine-furan springs show these bistable phenomena. In this article we introduce a coarse-grained model of such springs and investigate the possibility to observe the bistability effects, such as spontaneous vibrations. Our findings allow the studies of short pyridine-furan springs at large time and length scales and open field for further computational research of large bistable systems.
\end{abstract}


\section{Introduction}
\label{sec:I}
\begin{figure}
  \centering
    \includegraphics[width=0.5\textwidth]{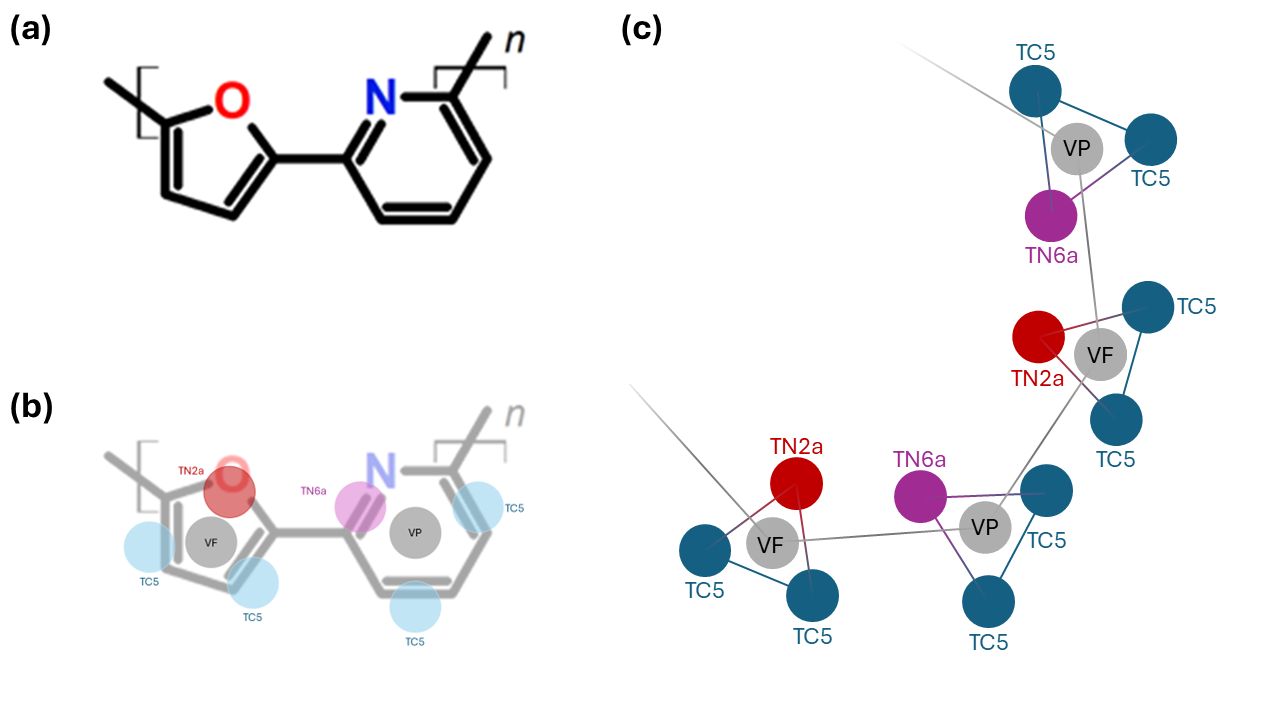}
  \caption{Pyridine-furan (PF) spring with five monomer units (oligo-PF-$5$ spring): (a) Chemical structure of a pyridine-furan monomer unit with heterocyclic rings in cis-configuration. (b) Mapping scheme of oligo-PF spring. (c) Two coarse-grained monomer units of oligo-PF spring.}
  \label{fig:fig1}
\end{figure}

This article introduces a coarse-grained model of spiral-like compound, pyridine-furan (PF), which consists of 6- and 5-member heterocycles (consult Figure \ref{fig:fig1}(a)), containing nitrogen and oxygen respectively\cite{ALANJONES19968707,jones1997extended}. This copolymer tends to form a helical shape stabilized by the interaction of $\pi$-electrons between neighboring turns, as shown by quantum calculations\cite{sahu2015}. Our previous modeling reveals that oligo-PF springs exhibit bistable dynamics of Duffing oscillators, accompanied by spontaneous vibrations and stochastic resonance activated by thermal fluctuations\cite{avetisov2021short,astakhov_spontaneous_2024}. Moreover, dynamical synchronization of two or several oligo-PF springs is also possible\cite{markina_spontaneous_2023}. Based on these findings, oligo-PF springs may serve as perspective candidates for the role of building elements in the design of nanoscale devices whose dynamics resemble classic bistable mechanical systems. The range of such devices may be from energy harvesters\cite{li_energy_2014,kim_harvesting_2015,ackerman_anomalous_2016,thibado_fluctuation-induced_2020} to switches and logic gates\cite{mi6081046,C5SC02317C,C7CS00491E,benda_substrate-dependent_2019,berselli_robust_2021,NICOLI2021213589} also including mechano-electrical converters\cite{dutreix_two-level_2020,cao_bistable_2021} and sensors and actuators\cite{zhang_molecular_2018,shu_stimuli-responsive_2020,lemme_nanoelectromechanical_2020,shi_driving_2020,aprahamian_future_2020,chi_bistable_2022}. Moreover, bistable nanodevices may also aid in validation of stochastic thermodynamics\cite{evans_fluctuation_2002,seifert_stochastic_2012,ciliberto_experiments_2017,horowitz_thermodynamic_2020}, which at the moment undergoes expansion to include molecular nanomachines\cite{ciliberto_experiments_2017,wang_experimental_2002,jop_work_2008,astumian_stochastic_2018,vroylandt_efficiency_2020}.

Nevertheless, a raise in the number of synchronized oligo-PF springs inevitably leads to the increase of the simulation box and subsequently to the slower simulation times. Moreover, our theoretical findings suggest an increase of lifetime of oligo-PF springs in each bistable state. Thus if we desire to simulate synchronization of a large number of oligo-PF springs, we have to compute more time steps while each time step is calculated slower, making all-atom modeling almost impossible.

An easy and elegant way to overcome this issue is coarse-graining (CG). There are two approaches to the CG modeling: “bottom-up,” which concentrates on reproduction of the chemical structure of the compound, and “top-down,” which focuses more on representation of its macroscopic properties. The main disadvantage of "bottom-up" models, despite their accuracy, is their lack of transferability; that is,  re-parametrization is required whenever modeling conditions change. Also, they often use complex potential forms that slow down calculations. At the same time, "top-down" approaches use simple potential forms and usually make use of the idea of pre-parameterized building blocks, making them easily transferable and computationally cheap.

Among various "top-down" CG force fields stands out Martini (which usually deciphers as MARrink Toolkit INItiative) that uses water/oil partitioning free energy to parameterize Lennard-Jones (LJ) potentials\cite{marrink_coarse_2004,marrink_martini_2007}. The keystone of the Martini force field is several discrete interaction levels between a limited number of building blocks. First introduced for lipids\cite{marrink_coarse_2004}, later it has been updated to include more compounds in its second version\cite{marrink_martini_2007} including proteins\cite{monticelli_martini_2008}, carbohydrates\cite{lopez_martini_2009}, DNA\cite{uusitalo_martini_2015} and various other compounds. Recently, a new version, coined Martini 3, with improved accuracy has been released\cite{souza_martini_2021,alessandri_martini_2022}. The main difference is the introduction of so-called "tiny" particles; this improvement plays a significant role for more accurate description of ring structures. Thus, picking Martini 3 for oligo-PF springs seems to be a straightforward choice.

In this paper we introduce a coarse-grained model of the oligo-PF spring within the latest version of Martini force filed. The model captures the bistable dynamic effects of these oligo-PF springs with thermally-activated spontaneous vibrations and stochastic resonance. Since both effects were already explored by us using all-atom molecular dynamics\cite{avetisov2021short,astakhov_spontaneous_2024}, we present the results in a manner that allows direct comparison with a previously published material.

\section{Materials and Methods}
\label{sec:MM}

\subsection{Non-bonded parametrization}\label{sec:MM1}
Martini $3$ incorporates four main types of beads: charged (Q), polar (P), non-polar (N) and apolar (C). There are also special beads for water (W), divalent ions (D) and groups containing halogen atoms (X). Most bead types except for W-bead type and D-bead type are divided in subtypes. For the simplicity, the LJ interaction levels $\epsilon$ are discretized; there are 22 possible interaction levels. Sizes of the beads $\sigma$ are also discretized in Martini 3: there are regular (R, $\sigma=\SI{0.47}{\nano\metre}$), small (S, $\sigma=\SI{0.41}{\nano\metre}$) and tiny (T, $\sigma=\SI{0.34}{\nano\metre}$) beads. These beads are intended to be used for $4$-to-$1$, $3$-to-$1$ and $2$-to-$1$ mappings respectively. Both pyridine and furan are already parameterized within Martini $3$ force field\cite{alessandri_martini_2022}, and more thorough discussion will follow in the next subsection \ref{sec:MM2}. Water is represented by a regular W bead.

\subsection{Bonded parametrization}\label{sec:MM2}
Aromatic ring structures are well-described by three T-beads: $($TC$5)_{2}$-TN$2a$ is a representation for furan and $($TC$5)_{2}$-TN$6a$ - for pyridine (TN$2a$ bead includes oxygen, TN$6a$ - nitrogen; consult Figure \ref{fig:fig1}(b-c) for more details). It is important to emphasize that in furan TN$2a$ bead shares $-CH-$ groups with neighboring TC$5$ beads, while in pyridine beads does not share atomic groups (see Figure \ref{fig:fig1}(b))\cite{alessandri_martini_2022}. \citet{alessandri_martini_2022} determined bonded parameters for standalone pyridine and furan, but bond lengths will differ in compound. Here, an important difference from the previous issue of Martini is that center of each bead corresponds to the center of geometry (COG) of underlying molecular structure rather that center of mass (COM). This is especially relevant for tiny beads.

To obtain bond potentials within the oligo-PF-$5$ molecule we adopt a strategy for long chains of rigid fragments called "Divide and Conquer"\cite{alessandri_martini_2022,alessandri_bulk_2017}. Within this strategy we introduce a virtual dummy site (mass-less and not interacting via non-bonded interactions) at the center of geometry of each pyridine or furan ring. Then, we connect them with an harmonic bond and exclude the nonbonded interactions between the two rings (see Figure \ref{fig:fig1}(c)). Valence angle potentials, such as $V_i$-$V_j$-$V_k$, $B_{i1}$-$V_i$-$V_j$ or $B_{i2}$-$B_{i3}$-$V_j$ where $V_i$, $V_j$ and $V_k$ are the $i$-th, $j$-th and $k$-th virtual sites and $B_{i1}$, $B_{i2}$, $B_{i3}$ are the beads of the $i$-th ring, may also be introduced (here for $B_{i1}$ it is convenient to take the bead describing oxygen/nitrogen; consult Figure \ref{fig:fig1}(c) for more details). Last, a dihedral potential $B_{i1}$-$V_i$-$V_j$-$B_{j1}$ can be directly applied in such a construction.

\subsection{Computational details}\label{sec:MM3}
For modeling we use Gromacs $2023$\cite{abr2015} with the number of particles, volume and temperature maintained constant (NVT) ensemble. The temperature is set to $\SI{298}{\kelvin}$ and maintained by a velocity-rescale thermostat\cite{bussi2007canonical} with $\SI{1.0}{\pico\second}$ coupling time. For parametrization we perform $\SI{10}{\nano\second}$ atomistic simulations of oligo-PF-$5$ spring using OPLS-AA\cite{kam2001} force field parameters for the oligomer, and the SPC/E model\cite{ber1987} for water. After obtaining atomistic trajectories, we calculate reference probability distributions for valence bonds and valence angles and get initial CG potentials, which are used for a primary $\SI{25}{\nano\second}$ CG simulation run; then we compare CG probability distributions with reference ones. We keep changing CG potentials until reasonable convergence between probability distributions is obtained. For CG calculations we use straight cutoff LJ potential that turns to zero at $\SI{1.1}{\nano\metre}$\cite{de_jong_martini_2016}.

To study the bistable dynamics of the oligo-PF-$5$, we fixed one end of the spring, while a pulling force was applied along the axis of the spring to another end of the spring. The distance (denoted $R_{e}$) between the ends of the oligo-PF-$5$ spring was considered a collective variable describing the long-term dynamics of the spring. Bistability of the oligo-PF-$5$ spring was specified in the agreement with two well-reproduced states of the spring with the end-to-end distances equal to $R_{e}\sim \SI{0.4}{\nano\metre}$ and $R_{e}\sim \SI{0.75}{\nano\metre}$. These states are referred to as the squeezed and the stress-strain states, respectively.

\section{Results}\label{sec:R}
\subsection{Non-bonded and bonded interactions}\label{sec:R1}
As mentioned in subsections \ref{sec:MM1} and \ref{sec:MM2}, both pyridine and furan are already parameterized in Martini $3$, but bonded potentials in a conjugated compound differ from those in a single heterocycle. To obtain these potentials, we first perform $\SI{10}{\nano\second}$ atomistic simulations of oligo-PF-$5$ using OPLS-AA force field with NVT ensemble. Temperature is maintained by velocity-rescale thermostat with $\SI{1.0}{\pico\second}$ coupling time. Then, we calculate probability distributions for all possible bonds and valence angles and thus get primary CG potentials which are then used for a trial run. The probabilities got from atomistic trajectory serve as target ones. We keep modifying CG potentials until the centers and widths of the distributions correspond the target ones. The obtained values are given in Table \ref{Tab1}

\begin{table}
\caption{Intramolecular potentials of oligo-PF-$5$\label{Tab1}. Here, subscripts $_{\text{F}}$ and $_{\text{P}}$ denote belonging to furan and pyridine. An asterisk ($*$) denotes a bead from neighboring ring}
\centering
\begin{tabular}{ccc|ccc}
\hline
{Valence bond}&{$r_{0}, \SI{}{\nano\metre} $}&{$k_{\text{b}}, \SI{}{\kilo\joule\per\mole\per\nano\metre\squared}$}&{Valence angle}&{$\theta_{0}$}&{$k_{\theta}, \SI{}{\kilo\joule\per\mole}$}\\
\hline
{TN$2a$-TC$5_{\text{F}}$}&0.1765&rigid&{V$_{\text{F}}$-V$_{\text{P}}$-V$_{\text{F}}$}&120&1500\\
{TC$5_{\text{F}}$-TC$5_{\text{F}}$ }&0.2125&rigid&{V$_{\text{P}}$-V$_{\text{F}}$-V$_{\text{P}}$}&132&1500\\
{TN$6a$-TC$5_{\text{P}}$}&0.25&rigid&{TN$2a$-V$_{\text{F}}$-($*$)V$_{\text{P}}$}&67&1000\\
{TC$5_{\text{P}}$-TC$5_{\text{P}}$}&0.255&rigid&{TN$6a$-V$_{\text{P}}$-($*$)V$_{\text{F}}$}&80&1000\\
{V$_{\text{F}}$-V$_{\text{P}}$}&0.438&50000&{V$_{\text{F}}$-($*$)V$_{\text{P}}$-($*$)TN$6a$}&40&1000\\
&&&{TC$5_{\text{P}}$-TC$5_{\text{P}}$-($*$)V$_{\text{F}}$}&161&500\\
&&&{V$_{\text{F}}$-($*$)TC$5_{\text{P}}$-($*$)TC$5_{\text{P}}$}&110&500\\
&&&{TC$5_{\text{F}}$-TC$5_{\text{F}}$-($*$)V$_{\text{P}}$}&145&500\\
\hline
\end{tabular}
\end{table}
Since in pyridine atomic groups are not shared between beads, an assymetry is observed in certain pairs of valence angles: TN$6a$-V$_{\text{P}}$-($*$)V$_{\text{F}}$ and V$_{\text{F}}$-($*$)V$_{\text{P}}$-($*$)TN$6a$ or TC$5_{\text{P}}$-TC$5_{\text{P}}$-($*$)V$_{\text{F}}$ and V$_{\text{F}}$-($*$)TC$5_{\text{P}}$-($*$)TC$5_{\text{P}}$ (consult Figure \ref{fig:fig1}(b) for more details). Also, a dihedral potential TN$6a$-V$_{\text{P}}$-($*$)V$_{\text{F}}$-($*$)TN$2a$ is directly taken from OPLS-AA force field.

\subsection{Spontaneous vibrations}\label{sec:R2}
\begin{figure}
  \centering
  \includegraphics[width=\textwidth]{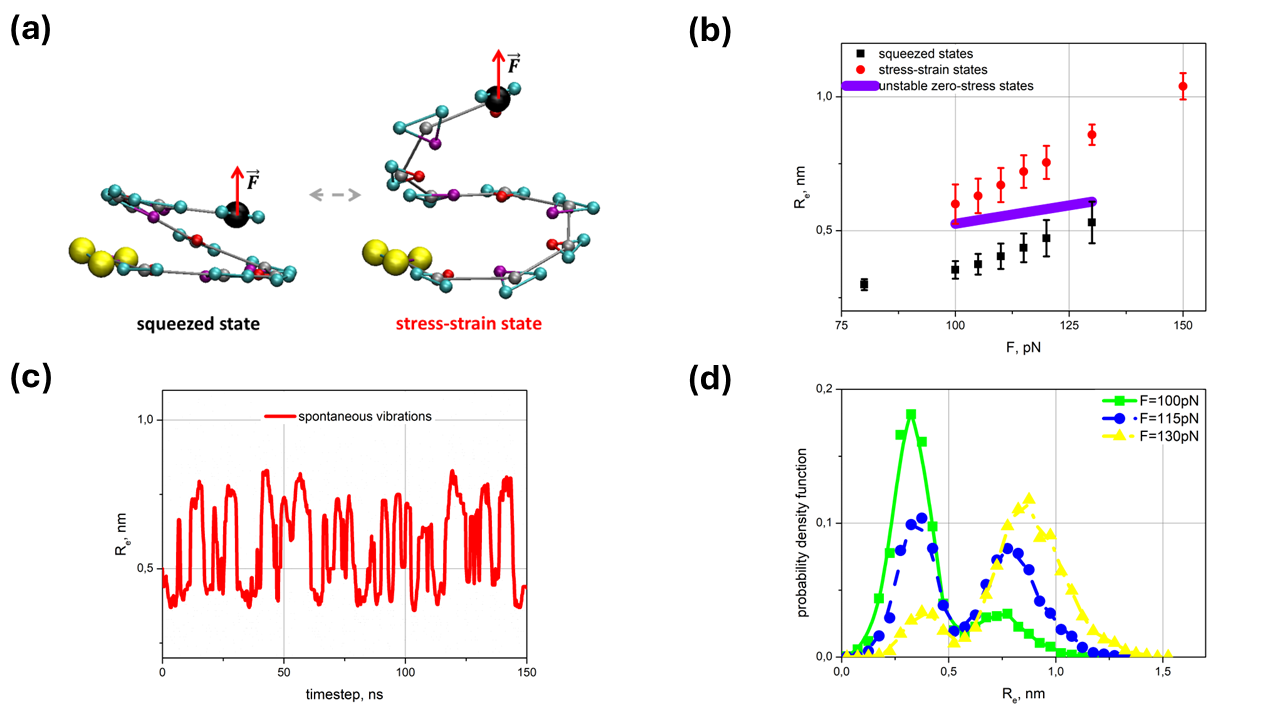}
  \caption{(a) Coarse-grained model of the oligo-PF-$5$ system with a pulling force. The squeezed and the stress–strain states of the spring are shown on the left and right, respectively. The yellow spheres at the lower end of the spring indicate the fixation of the pyridine ring by rigid harmonic force. The pulling force, $F$, is applied to the top end of the spring. (b) The state diagram shows a linear elasticity of oligo-PF-$5$ spring up to  $F\approx \SI{100}{\pico\newton}$ and bistability of the spring in the region from $F\approx\SI{100}{}- \SI{130}{\pico\newton}$; (c) Spontaneous vibrations of the oligo-PF-$5$ spring at  $F\approx \SI{115}{\pico\newton}$; (d) Evolution of the probability density for the squeezed and stress–strain states when pulling force surpasses the critical value.}
  \label{fig:fig2}
\end{figure}

To study the behavior of the oligo-PF-$5$ spring under tension, we used the same approach as for the all-atom model.\cite{avetisov2021short}. Having equilibrated the oligo-PF-$5$ spring at $\SI{298}{\kelvin}$ fixing one end of it we applied a force $\vec{F}$ along the spring axis to pull the other end. In conditions of weak tension, the initial state of the spring remained stable; but as soon as the tension force reached a critical value (approximately $F_c = \SI{100}{\pico\newton}$ ), the spring began to spontaneously vibrate between squeezed and stress–strain states.The average end-to-end distances of the spring in these states differ by approximately $\SI{0.35}{\nano\metre}$,  which is consistent with the results of $ab$ $initio$ calculations\cite{sahu2015} and atomistic modeling. Figure \ref{fig:fig2}(a) displays coarse-grained level snapshots of these two states.

The evolution of the statistics of visits to the squeezed and stress-strain states as the pulling force exceeds the critical point $F_{c}$ is shown on Figures \ref{fig:fig2}(b,d). In the range of force $F$ from $110$ to $\SI{120}{\pico\newton}$, the squeezed and stress-strain states occur almost with the same frequency. The average lifetime of the states in the mode of spontaneous vibrations in the bistability region varies from $\tau = 1$ to $\SI{30}{\nano\second}$, depending on the tensile force. In the symmetrical bistability region, neither the squeezed state nor the stress-strain state prevails, resulting in the average lifetime of these two states being approximately equal to $\tau = \SI{3.6}{\nano\second}$. Figure \ref{fig:fig2}(c) shows a typical trajectory of long-term dynamics $R_{e}(t)$ of the oligo-PF-$5$ spring in the symmetrical bistability region. It can be seen that spontaneous vibrations of the spring occur without any additional random influences. These vibrations are activated only by fluctuations of the thermal reservoir. Outside the bistability region, trajectories without vibrations prevail.

\subsection{Stochastic resonance}\label{sec:R3}
\begin{figure}
  \centering
  \includegraphics[width=\textwidth]{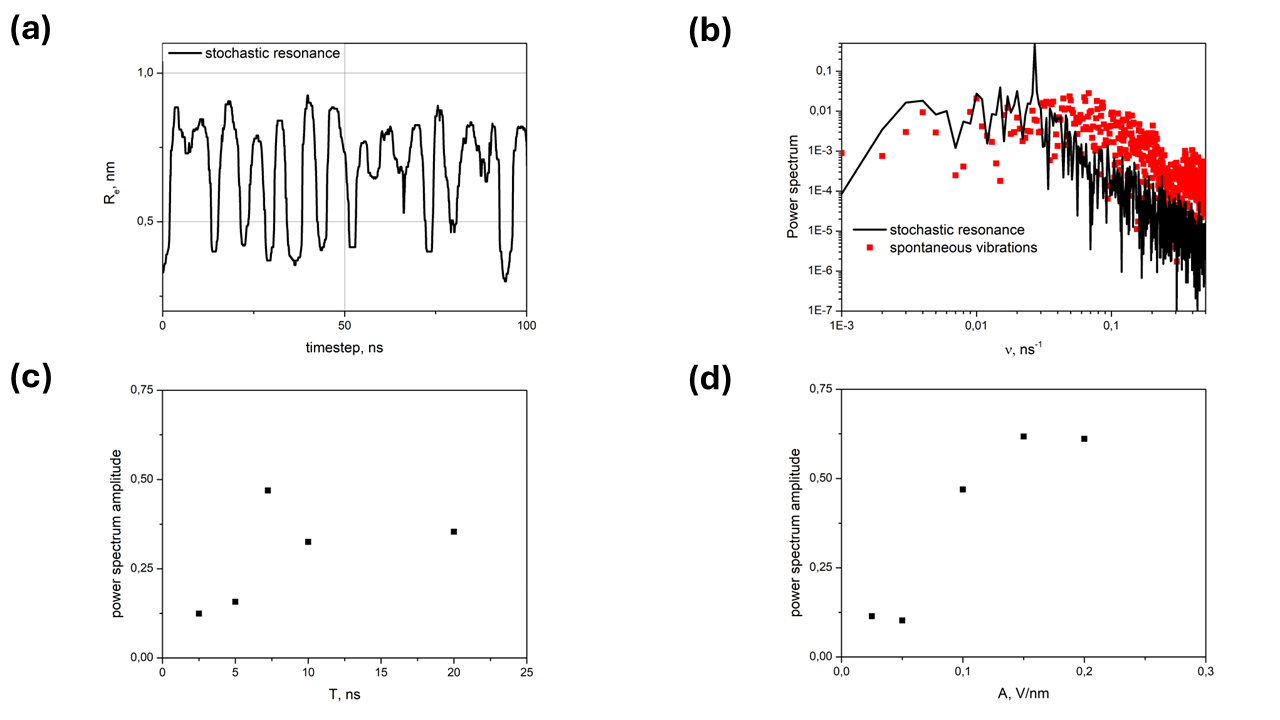}
  \caption{Stochastic resonance of the oligo-PF-$5$ induces by an oscillating field $E = E_{0} \cos (2 \pi \nu t)=E_{0} \cos (\nicefrac{2 \pi t}{T})$: (a) The dynamic trajectory at $F = \SI{115}{\pico\newton}$, $T = \SI{7.5}{\nano\second}$, and $E_{0} = \SI{0.2}{\volt\per\nano\metre}$; (b) Power spectra of spontaneous vibrations (red curve) and stochastic resonance (black curve); (c) The dependence of the main resonance peak amplitude on the period $T$ of oscillating field ($E_{0}=\SI{0.1}{\volt\per\nano\metre}$); (d) The dependence of the main resonance peak amplitude on $E_{0}$ ($T_{0}=\SI{7.2}{\nano\second}$).}
  \label{fig:fig3}
\end{figure}

To study the stochastic resonance mode of the oligo-PF-5 spring, we added a weak periodic force to the pulling end of the spring. For this purpose, we simulated the effect of an oscillating electric field $E = E_0 \cos(2\pi \nu t)$ on a unit charge placed at the pulling end of the spring. Since CG particles are uncharged, such an addition should not affect the dynamics of the system.Typical dynamic trajectory of the oligo-PF-5 spring in the stochastic resonance mode is shown in Figure \ref{fig:fig3}(a) while the power spectra of vibrations in both spontaneous vibration modes and stochastic resonance modes are displayed in Figure \ref{fig:fig3}(b).

The main resonance peak was detected at the frequency $\nu=(2\tau)^{-1}$, which corresponds to the situation when the period of the applied oscillating field is equal to twice the average lifetime of a state in the mode of spontaneous vibrations, as predicted by the theory of stochastic resonance\cite{gammaitoni_stochastic_1998,wellens_stochastic_2004}. We tested various oscillating fields to find the conditions for stochastic resonance, and the results are shown in Figure \ref{fig:fig3}(c,d). The amplitude of the main resonance peak increases with the amplitude of the oscillating field up to a value of $E_{0}\approx\SI{0.1}{}-\SI{0.15}{\volt\per\nano\metre}$; further oscillating field amplitude's increase does not affect the resonance peak. Considering that high amplitudes of the oscillating field may cause forced oscillations, we set $E_{0}=\SI{0.1}{\volt\per\nano\metre}$ as the upper limit below which the stochastic resonance occurs. It worth noting that outside the symmetric bistability region, the lifetimes of the squeezed and the stress–strain states became so different that the average lifetime ceases to be a reliable indicator of the resonance frequency, and thus the resonant response study within this work is limited to the region of symmetric bistability at $F=\SI{115}{\pico\newton}$.
\section{Discussion}\label{sec:D}
The main finding of this work is that subtle bistable effects, such as spontaneous vibrations and stochastic resonance, can be observed using coarse-grained molecular dynamics. Indeed, our previous studies suggest that spontaneous vibrations arise due to $\pi$-stacking between aromatic groups located on the adjacent turns of the spring, which along with the stiffness of the molecular backbone produces a two-well potential (see \citet{avetisov2021short} for more information). One can speculate how can $\pi$-stacking be addressed in molecular dynamics (authors may suppose that it somehow emerges due to a peculiar interplay between Lennard-Jones forces and Coulomb interaction of the partial charges of the atoms of the rings, but such question is very far out of the scope of the current work), but in the present Martini model adjacent turns of the spring interact exclusively via LJ potential. In the meantime, chain stiffness is kept by bonding between virtual dummy particles; both factors seem strong arguments against observing bistability. Thus, such emerging noticeably strengthens the credibility of Martini force field and provides for its wider usability for the molecular modeling field.

Still, there is a certain difference between spontaneous vibrations observed in the present work and in our previous atomistic modeling\cite{avetisov2021short}. In CG modeling, SV occur at smaller forces: \SI{100}{}--\SI{120}{\pico\newton} compared with \SI{270}{}--\SI{290}{\pico\newton} for all-atom modeling. Nevertheless, we do not think it should be considered as significant disadvantage of the model: the fact that the mean lifetimes in the states \SI{1}{}--\SI{30}{\nano\second} correspond to those obtained in fully atomic simulation and that the widths of the force regions coincide seems to be more important.

For the present work we have taken pre-parameterized pyridine and furan\cite{alessandri_martini_2022} and as a consequence have determined only bonded potentials. Since TN$2a$ bead in furan shares -$CH$- group with adjacent TC$5$ beads while in pyridine 
beads does not share molecular groups with their neighbors, we are obliged to introduce differing valence angle potentials (for instance, TN$6a$-V$_{\text{P}}$-($*$)V$_{\text{F}}$ and V$_{\text{F}}$-($*$)V$_{\text{P}}$-($*$)TN$6a$) due to asymmetry of the resulting construction. This could be overridden by a slight re-parametrization of pyridine: if TN$6a$/TC$5$ beads representing pyridine shared -$CH$- groups, the construction would be symmetrical and valence angle potentials would be the same. We plan to investigate more possible mappings in our further work.

In our previous work concerning Martini\cite{muratov_modeling_2021} we noticed the ambiguity in relation to bead sizes. Fortunately, Martini $3$ has addressed the issues we mentioned by introducing tiny beads and strict definition when small and tiny beads are to be used\cite{souza_martini_2021}. The latter preserves flexibility while implying more accuracy.

\section{Conclusion}\label{sec:C}
We have performed coarse-grained simulations of oligo-PF-$5$ springs subject to stretching and observed thermally-activated spontaneous vibrations and stochastic resonance. We explored the conditions under which these effects are present, indicated symmetrical bistability and determined mean lifetime of the states. The results are in good accordance with atomistic simulations. Our findings approve Martini force field as capable tool for molecular modeling. Its combined flexibility and accuracy allows its applying for the modeling of molecular machines since it is able to capture the necessary time- and lengthscales. 

\section*{Acknowledgments}
Authors thank Alexey Astakhov, Vladimir Bochenkov, Maria Frolkina, Vladislav Petrovskii, Anastasia Markina and Alexander Valov for helpful discussions.

\bibliographystyle{unsrtnat}  

\end{document}